\documentclass[
    ,final            
  ]
  {aipproc}

\layoutstyle{6x9}

\begin{document}

\hfill JLAB-THY-09-1019

\title{Binding of $D$, $\overline{D}$ and $J/\Psi$ mesons in nuclei}

\classification{21.85.+d,21.65Jk,21.65.Qr,24.85.+P}

\keywords{Meson-nuclear bound state, $D,\overline{D}$ and $J/\Psi$ in nuclei,
Quark-meson coupling model}

\author{K. Tsushima}{
  address={
  EBAC in Theory Center, Jefferson Lab, 12000 Jefferson Ave. Newport News, VA 23606, USA}
}

\begin{abstract}
We estimate the $D$, $\overline{D}$ and $J/\Psi$ meson potentials
in nuclei. $J/\Psi$-nuclear potential (self-energy)
is calculated based on the color-singlet mechanism, by the $DD$ and $DD^*$ meson loops.
This consistently includes the in-medium properties of $D$ and $D^*$ mesons.
The potential obtained for the $J/\Psi$ meson is attractive, and it is complementary
to the attraction obtained from the color-octet gluon-based mechanism.
We predict that the $D^-$ and $J/\Psi$ mesons will be bound in nuclei under proper
conditions.

\end{abstract}

\maketitle


\subsection{Introduction}

The properties of charmed mesons and charmonium in a nuclear medium (nuclei)
are still very little known. However, future experimental facilities such as JLab
(after 12 GeV upgrade of CEABAF), J-PARC, and FAIR will make it possible to explore
the in-medium properties of these mesons.
Among them, below we focus on the $D$, $\overline{D}$ and $J/\Psi$ mesons.

\begin{figure}[htb]
\includegraphics[height=.35\textheight,angle=-90]{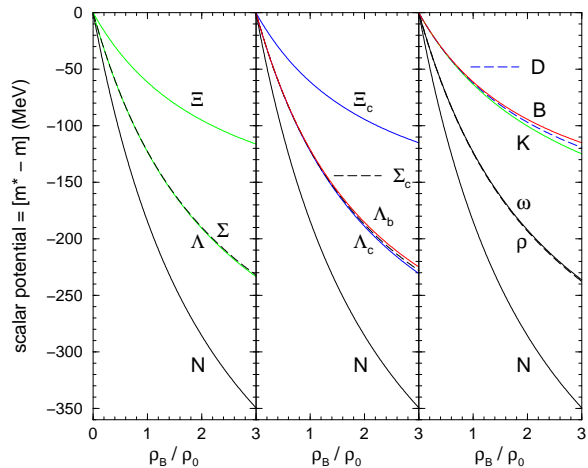}
\caption{Scalar potentials (in-medium mass minus free mass)
in symmetric nuclear matter~\cite{QMCbcmatter}.}
\label{spot}
\end{figure}

To study the in-medium properties of $D$, $\overline{D}$ and $J/\Psi$ mesons,
we rely on the quark-meson coupling (QMC) model~\cite{Guichon,QMCreview}.
The QMC model is a quark-based, relativistic mean field model of nuclear
matter and nuclei~\cite{Guichon,QMCreview}. Relativistically moving confined light quarks in
the nucleon bags self-consistently interact directly with the scalar-isoscalar $\sigma$,
vector-isoscalar $\omega$, and vector-isovector $\rho$ mean fields generated by the
light quarks in the (other) nucleons. These meson mean fields are responsible for the nuclear binding.
The direct interaction between the light quarks and the scalar $\sigma$ field
is the key of the model, which induces the {\it scalar polarizability} at the nucleon level,
and generates the nonlinear scalar potential (effective nucleon mass), or
the density ($\sigma$-filed) dependent $\sigma$-nucleon coupling.
This gives a novel, new saturation mechanism for nuclear matter.
The model has opened tremendous opportunities for the studies of finite nuclei
and hadrons properties in a nuclear medium (nuclei), based on the quark degrees of freedom.
Many successful applications of the model can be found in Ref.~\cite{QMCreview}.

In the QMC model, since the couplings between the light quark and the $\sigma$, $\omega$,
and $\rho$ fields are the same for all the light quarks in the hadrons irrespective of the hadron
species, the model can treat them systematically in the nuclear medium.
In particular, the scalar potentials (in-medium mass minus free mass) for
hadrons have turned out to be proportional to the number of
light quarks in each hadron --- the light quark number counting rule~\cite{QMCbcmatter}.
This is demonstrated in Fig.~\ref{spot}.

\subsection{$D$ and $\overline{D}$ mesons in symmetric nuclear matter and nuclei}

\begin{figure}[htb]
  \includegraphics[height=.25\textheight]{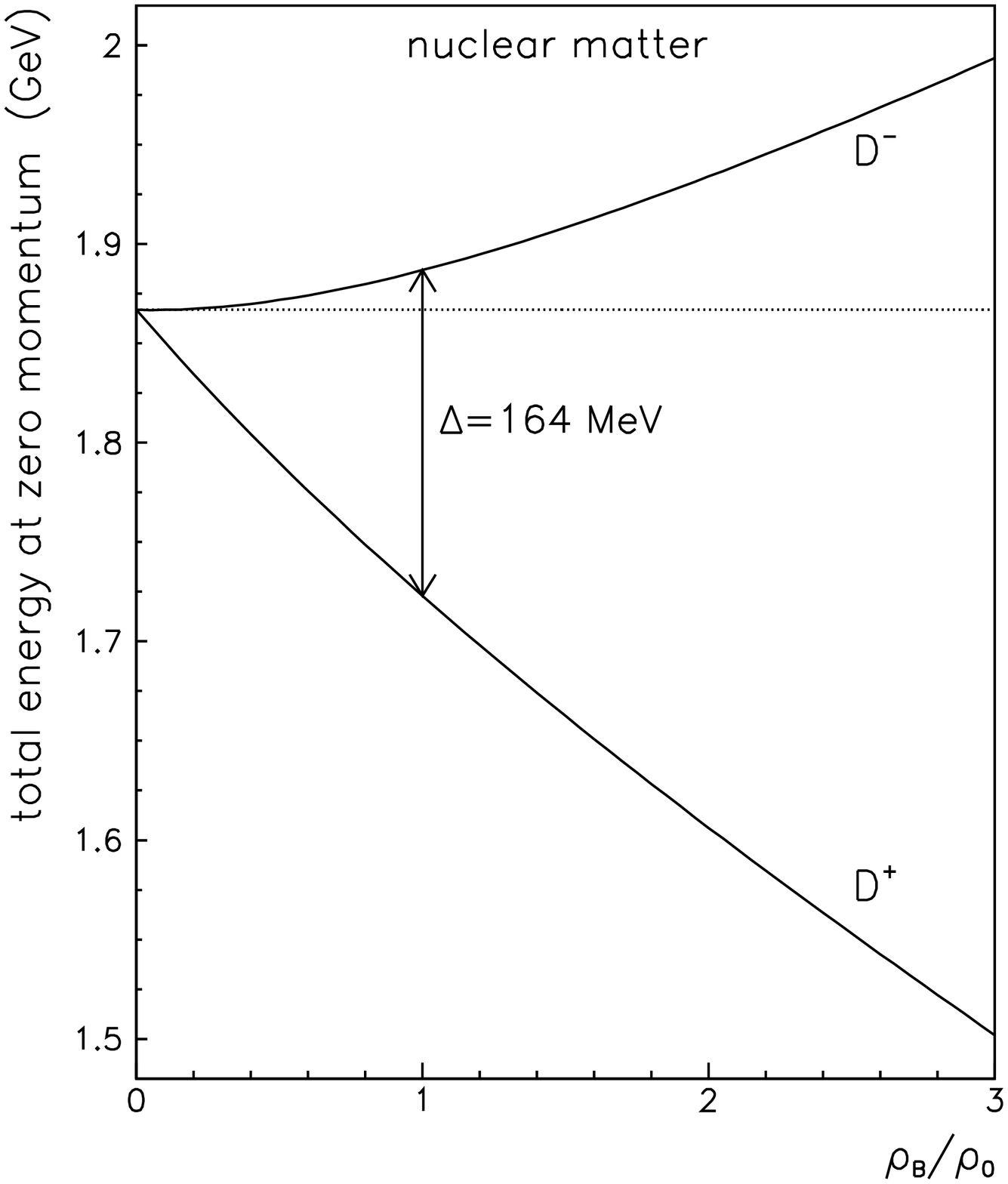}
  \includegraphics[height=.25\textheight]{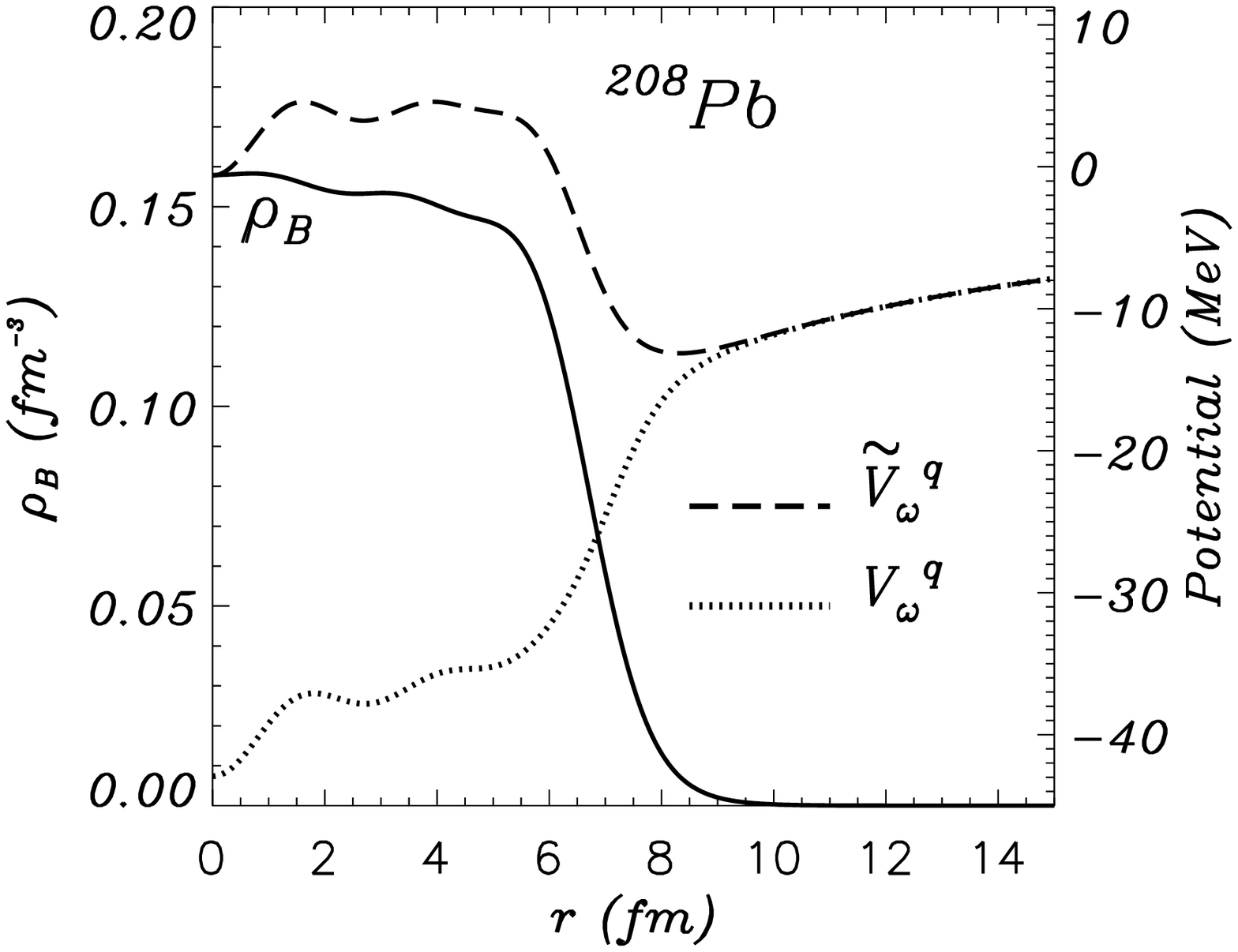}
  \caption{$D$ and $\overline{D}$ meson potentials in symmetric nuclear matter
(left panel)~\cite{Alexcharm}, and total potential of the $D^-$ in a Pb nucleus including the Coulomb
potential (right panel)~\cite{QMCcharm}. Note that the $\omega$ vector potential (aside from the signs)
for the $D$ and $\overline{D}$ in the left panel is $\tilde{V}^q_\omega = 1.96 V^q_\omega$,
the same as that used for kaon~\cite{kpot}.
}
\label{dpot}
\end{figure}

First, we recall the in-medium properties of $D$ and $\overline{D}$ mesons
studied in QMC~\cite{Alexcharm,QMCcharm}.
In Fig.~\ref{dpot} (left panel) we show the total (scalar plus vector) potentials
for the $D$ and $\overline{D}$ mesons in symmetric nuclear matter.
To calculate the $J/\Psi$ potential (self-energy) in (symmetric) nuclear matter by the $D$ and $D^*$
meson loops, in-medium properties of the $D$ and $D^*$ must be consistently included.
This will be done in next section.

In Ref.~\cite{QMCcharm} we studied possibilities of forming meson-nuclear bound states
for the $D$ and $\overline{D}$ mesons in a Pb nucleus.
The single-particle energies obtained are listed in Table~\ref{energy}.
Among them, the $D^-$ meson is promising to explore experimentally.
Since its quark content is $D^- = \bar{c}d$, the width is expected to be narrow
due to no light-quark annihilation channels. Furthermore, because of the attractive Coulomb
potential, there is always a chance to form the Coulomb bound state irrespective
of the details in the strong interaction potential inside the Pb nucleus.
We show in Fig.~\ref{dpot} (right panel) the total potential calculated for
the $D^-$ meson in the Pb nucleus including the Coulomb potential,
for the two cases of the vector potentials,
$V^q_\omega$ and $\tilde{V}^q_\omega = 1.96V^q_\omega$.

\begin{table}[htb]
\caption{
$D^-$, $\overline{D^0}$ and $D^0$ meson single-particle energies (in MeV)
in $^{208}$Pb for different potentials.
The widths for the mesons are all set to zero, both
in free space and inside $^{208}$Pb. Note that the $D^0$ bound state
energies calculated with $\tilde{V}^q_\omega = 1.96 V^q_\omega$ will be much larger than
those calculated with $V^q_\omega$ in absolute value. (From Ref.~\cite{QMCcharm}.)
}
\label{energy}
\begin{tabular}[t]{lccc|ccc}
\hline
state  &$D^- (\tilde{V}^q_\omega)$ &$D^- (V^q_\omega)$
&$D^- (V^q_\omega$, no Coulomb) &$\overline{D^0} (\tilde{V}^q_\omega)$
&$\overline{D^0} (V^q_\omega)$ &$D^0 (V^q_\omega)$ \\
\hline
                         1s &-10.6 &-35.2 &-11.2 &unbound &-25.4 &-96.2\\
                         1p &-10.2 &-32.1 &-10.0 &unbound &-23.1 &-93.0\\
                         2s & -7.7 &-30.0 & -6.6 &unbound &-19.7 &-88.5\\
\end{tabular}
\end{table}

\subsection{$J/\Psi$ meson potential in symmetric nuclear matter}

The $J/\Psi$ potential in nuclear matter (nuclei) has been estimated mainly based on
the gluon induced mechanisms --- based on an effective filed theory~\cite{Luke},
QCD Stark effect~\cite{Lee}, and chromo-polarizability~\cite{Voloshin}.
In the following, we study the $J/\Psi$ potential (self-energy) in symmetric nuclear matter by
the color-singlet mechanism, the $DD$ and $DD^*$ meson loops, where the relevant vertexes
are $J/\Psi$-$D$-$D$ and $J/\Psi$-$D$-$D^*$, respectively.
An analysis combined with the $D^*D^*$ meson loop contribution
will be reported elsewhere~\cite{GKT}.

\begin{figure}[htb]
\includegraphics[height=.4\textheight,angle=-90]{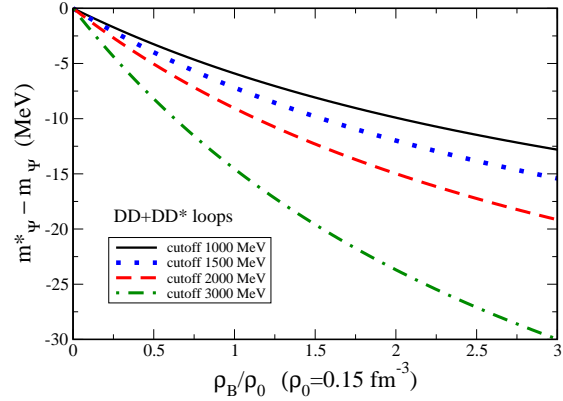}
\caption{$J/\Psi$ potential in symmetric nuclear matter.
The solid, dotted, dashed, and dash-dotted curves correspond to the common cut-off values
in the dipole form factors attached to the $J/\Psi$-$D$-$D$ and $J/\Psi$-$D$-$D^*$ vertexes,
1000, 1500, 2000, and 3000 MeV, respectively.}
\label{psipot}
\end{figure}

To calculate the in-medium $J/\Psi$ self-energy by
the $DD$ and $DD^*$ meson loops, we need to include the in-medium properties
of $D$ and $D^*$ mesons consistently as mentioned in the previous section.
In addition, we introduce the dipole form factors~\cite{latvec} for
the $J/\Psi$-$D$-$D$ and $J/\Psi$-$D$-$D^*$ vertexes with the common cut-offs.
The coupling constants used are, $g_{J/\Psi DD} = g_{J/\Psi DD^*} = 7.7$.
The $J/\Psi$ potential calculated in symmetric nuclear matter
(in-medium mass minus free mass), $m^*_{\Psi} - m_{\Psi}$,
is shown in Fig.~\ref{psipot}.

We regard the results with the cut-off values 1500 and 2000 MeV as our predictions.
At normal nuclear matter density, these correspond to about $7$ and $9$ MeV
attractions, respectively.
Combined with the contribution from the color-octet gluon-based
attraction~\cite{Luke,Lee,Voloshin}, and an additional attraction expected
from the $D^*D^*$ loop, we conclude that the $J/\Psi$ meson will be bound
in nuclei under proper conditions.
In this case, the width is expected to be narrow.
The experimental search for the $J/\Psi$-nuclear bound states will be
possible at JLab after 12 GeV upgrade of CEABAF.

\subsection{Summary and Conclusion}

We have studied binding of $D$, $\overline{D}$ and $J/\Psi$ mesons
in nuclei based on the quark-meson coupling model.
Due to the attractive Coulomb potential, the $D^-$ meson will be inevitably bound in
a Pb nucleus with a narrow width. This does not depend on the details of
the strong interaction potential for the $D^-$ in a Pb nucleus.
For the $J/\Psi$-nuclear potential, we have estimated based on the
color-singlet mechanism, by the $DD$ and $DD^*$ meson loops, consistently
including the in-medium properties of the $D$ and $D^*$ mesons.
Combined with the color-octet gluon-based attraction, and an additional attraction
expected from the $D^*D^*$ loop,
we expect that the $J/\Psi$ meson will be bound in nuclei with narrow widths
under proper conditions.


\begin{theacknowledgments}
The author would like to thank Marvin L.~Marshak for his warm support before and after
the conference, and G. Krein and A. W. Thomas for the collaboration.
Notice: Authored by Jefferson Science Associates, LLC under U.S. DOE Contract No. DE-AC05-06OR23177.
The U.S. Government retains a non-exclusive, paid-up, irrevocable, world-wide license to publish
or reproduce this manuscript for U.S. Government purposes.
\end{theacknowledgments}



\bibliographystyle{aipproc}   


\begin{thebibliography}{9}

\bibitem{Guichon}
P.~A.~M. Guichon, \emph{Phys. Lett. B} \textbf{200}, 235 (1988).

\bibitem{QMCreview}
K. Saito, K. Tsushima, A.~W.~Thomas, \emph{Prog. Part. Nucl. Phys.} \textbf{58}, 1 (2007).

\bibitem{QMCbcmatter}
K.~Tsushima and F.~C.~Khanna, \emph{Phys. Lett. B} \textbf{552}, 138 (2003).

\bibitem{Alexcharm}
A.~Sibirtsev, K.~Tsushima, A.~W.~Thomas, \emph{Eur. Phys. J. A.} \textbf{6}, 351 (1999).

\bibitem{QMCcharm}
K.~Tsushima {\it et al.}, \emph{Phys. Rev. C} \textbf{59}, 2824 (1999).

\bibitem{kpot}
K.~Tsushima, K.~Saito, A.~W.~Thomas and S.~V.~Wright,
  \emph{Phys. Lett.  B} \textbf{429}, 239 (1998)
  [\emph{Erratum-ibid.  B} \textbf{436}, 453 (1998)].

\bibitem{Luke}
M.~Luke, A.~V.~Manohar, M.~J.~Savage, \emph{Phys. Lett. B} \textbf{288}, 355 (1992).

\bibitem{Lee}
S.~H.~. Lee and C.~M.~Ko, \emph{Phys. Rev. C} \textbf{67}, 038202 (2003).

\bibitem{Voloshin}
M. B. Voloshin, \emph{Prog. Part. Nucl. Phys.} \textbf{61}, 455 (2008).

\bibitem{GKT}
G.~Krein, K.~Tsushima, A.~W.~Thomas, in preparation.

\bibitem{latvec}
D.~B.~Leinweber, A.~W.~Thomas, K.~Tsushima and S.~V.~Wright,
\emph{Phys. Rev. D} \textbf{64}, 094502 (2001).


\end{thebibliography}

%


%
\end{document}